\documentclass[conference]{IEEEtran}
\usepackage[utf8x]{inputenc}

\usepackage{cite}

\ifCLASSINFOpdf
  \usepackage[pdftex]{graphicx}
  \graphicspath{{./figures/}}
\else
  \usepackage[dvips]{graphicx}
  \graphicspath{{./eps/}}
\fi
\usepackage{enumerate}

\usepackage{array}

\usepackage{tabularx}

\ifCLASSOPTIONcompsoc
 \usepackage[caption=false,font=normalsize,labelfont=sf,textfont=sf]{subfig}
\else
  \usepackage[caption=false,font=footnotesize]{subfig}
\fi
\usepackage{url}

\begin{document}
%
\title{Increasing Traffic Flows with DSRC Technology:
Field Trials and Performance Evaluation}

\author{\IEEEauthorblockN{R. Zhang \IEEEauthorrefmark{1}, F. Schmutz \IEEEauthorrefmark{2}, K. Gerard \IEEEauthorrefmark{2}, A. Pomini \IEEEauthorrefmark{2}, L. Basseto \IEEEauthorrefmark{2}, S.~B. Hassen \IEEEauthorrefmark{2},  A. Jaiprakash \IEEEauthorrefmark{1}, I. Ozgunes \IEEEauthorrefmark{1},\\ A. AlArifi \IEEEauthorrefmark{3}, H. Aldossary \IEEEauthorrefmark{3}, I. AlKurtass \IEEEauthorrefmark{3},  O. Talabay \IEEEauthorrefmark{3}, A. AlMohanna \IEEEauthorrefmark{3}, S. AlGhamisi \IEEEauthorrefmark{3}, A.~A. Biyabani \IEEEauthorrefmark{4},\\ K. Al-Ghoneim \IEEEauthorrefmark{4}, and O.~K. Tonguz \IEEEauthorrefmark{1}}\\
\IEEEauthorblockA{\IEEEauthorrefmark{1} Department of Electrical and Computer Engineering,
Carnegie Mellon University,
 Pittsburgh, PA 15213-3890, USA}

\IEEEauthorblockA{\IEEEauthorrefmark{2}School of Computer and Communication Sciences, Ecole Polytechnique Federale de Lausanne (EPFL), Lausanne, Switzerland}

\IEEEauthorblockA{\IEEEauthorrefmark{3}King Abdulaziz City for Science and Technology (KACST),  Riyadh, Saudi Arabia}

\IEEEauthorblockA{\IEEEauthorrefmark{4}Hawaz Inc.,  Riyadh, Saudi Arabia}}


%


\maketitle

\vspace{-0.5in}

\begin{abstract}
As traffic congestion becomes a huge problem for most developing and developed countries across the world, intelligent transportation systems (ITS) are becoming a hot topic that is attracting attention of researchers and the general public alike. In this paper, we demonstrate a specific implementation of an ITS system whereby traffic lights are actuated by DSRC radios installed in vehicles. More specifically, we report the  design of prototype of a DSRC-Actuated Traffic Lights (DSRC-ATL) system. It is shown that this system can reduce the travel time and commute time significantly, especially during rush hours. Furthermore, the results reported in this paper do not assume or require all vehicles to be equipped with DSCR radios.  Even with low penetration ratios, e.g., when only 20\% of all vehicles in a city are equipped with DSRC radios, the overall performance of the designed system is superior to the current traffic control systems.


\end{abstract}
\vspace{0.1in}
\emph{keywords: V2I communications, Dedicated Short-Range Communications, vehicular networks, intelligent traffic lights, intelligent transportation systems, vehicle sensing}

%
\IEEEpeerreviewmaketitle

\section{Introduction}
\footnote[1]{The research reported in this paper was funded by King Abdulaziz City of Science and Technology (KACST), Riyadh, Kingdom of Saudi Arabia}
Traffic congestion is a formidable problem for developing and developed countries across the world.
Unfortunately, to date no viable solutions have been reported that is effective and low cost. In this paper, we address this issue by showing how a communications-based traffic control scheme could provide a viable and low cost solution to this daunting problem.

Intelligent Transportation Systems (ITS) have played a significant role in reducing daily commute times and alleviating traffic congestions at intersections. Traditional ITS employ intelligent intersections that can detect vehicles by using loop detectors, magnetic detectors or cameras \cite{trafficdetector} and adapt the decision of traffic lights accordingly. Such solutions are very costly and therefore have not scaled well in the last three decades in most cities (in the US the number of traffic lights equipped with camera systems and/or loop detectors is less than 10\% of all traffic lights). This strategy is very expensice due to the high cost of devices, installation and maintenance.


Dedicated Short-Range Communications (DSRC) technology is a very attractive new technology. While it was initially designed for traffic safety applications, in this paper we show that it can be leveraged for traffic efficiency and traffic control applications as well.

In this paper, we report
  a prototype of a DSRC-Actuated Traffic Lights (DSRC-ATL) system that can significantly improve the throughput at an intersection and also the travel time or commute time.
Field trials and simulations  clearly show that the system is able to reduce waiting time of commuters at the intersections,
even when only a low percentage of vehicles are equipped with DSRC radios.

\vspace{-0.05in}
\section{Related Work}

In the past few decades, various adaptive traffic systems were developed and implemented in some cities  \cite{trafficlight}. Some of these traffic systems such as SCOOT \cite{robertson1991optimizing,hunt1982scoot}, SCATS \cite{lowrie1990scats}, are based on dynamic traffic coordination \cite{ luk1984two}, and can be viewed as a traffic-responsive version of TRANSYT \cite{robertson1900tansyt}. These systems optimize the offsets of traffic signals in the network, based on current traffic demand, and generate `green-wave' for major car flow.  Meanwhile, some other model-based systems have been proposed, including OPAC \cite{gartner1983opac}, RHODES\cite{mirchandani2001real}, PRODYN\cite{henry1983prodyn}. These systems use both the current traffic arrivals and the prediction of future arrivals and choose a signal phase planning which optimizes the objective functions. While these systems work efficiently, they do have significant disadvantages: the cost of these systems is generally high and they are hard to install and maintain. Considering SCATS, for example, the initial cost of the system is \$20,000 to \$30,000 per intersection, and \$28,800 per mile per year, not to mention that the installation will cost an extra \$20,000 per intersection \cite{cost}.   The cost is due to the fact that these systems use loop detectors and video cameras to detect vehicles.

As Dedicated Short-Range Communication (DSRC) radios are installed in new vehicles, a vehicle-to-infrastructure (V2I) communications based vehicle detection method becomes viable. This method has a lot of benefits that other detection methods can't provide. First, it is robust against brightness, illumination, and weather. Second, it is easy to implement and maintain, and third, it is low cost. This detection method also provides more detailed information such as a vehicle's type, location, speed, trajectory. Several systems has been proposed in the past few years using DSRC for traffic control \cite{gradinescu2007adaptive, cai2010adaptive, avzekar2014adaptive}. These systems, however, are based on the futuristic concept that most of the vehicles will be equipped with DSRC radios. Hence, they cannot be implemented in the short term since all the industry forecasts project that the penetration of DSRC technology will occur gradually as opposed to immediately. It might take several years before the penetration of DRSC technology reaches the level desired for such schemes to be practical and implementable. Therefore, most of these schemes are, while interesting and forward-looking, difficult to implement because of this partial penetration problem.

There are several other related prototype systems reported recently, such as Compass4D \cite{vreeswijk2014compass4d} , and Cooperative Vehicle-Infrastructure System (CVIS) \cite{kompfner2007cooperative} which implemented several warning messages to the vehicles in order to improve vehicles' safety as well as improve energy efficiency and reduce congestions. However, the main goal of CVIS is to post information to vehicles as opposed to promoting the idea of vehicles directly interacting with traffic lights for improving traffic efficiency (i.e., increasing traffic flows). Eco-move \cite{vreeswijk2010energy} is yet another interesting DSRC based project aiming at reducing overall fuel consumption. Its sub-system EcoGreenWave presents a similar concept of managing a traffic network in order to achieve a minimum amount of CO2
emission, number of stops and delay. However, Eco-move also assumes that all vehicles are equipped with DSRC radios and, as such, does not address the partial penetration issue. In other words, it is not clear if any of these prototype systems can be implemented if only a small portion of the vehicles on the road (such as 20-30\%) are equipped with DSRC radios. COLOMBO is another interesting project that focuses on low-penetration rate of DSRC-equipped vehicles \cite{bellavista2014implementing,krajzewicz2013colombo,bellavista2014v2x}.  The designed system aims at using the information provided by V2X technology, as well as other data obtained from cellular systems and WiFi direct connections by feeding this information to a traffic management system. While the idea of COLOMBO seems synergistic to the DSRC-actuated traffic control system reported in this paper, the approach reported here is much simpler in terms of protocol design, as the method reported in this paper doesn't require communications between intersections. Meanwhile, unlike the solution reported in this paper, COLOMBO cannot react to real-time traffic flows. This implies that DRSC-equipped vehicles will not get a better performance compared to unequipped vehicles. Thus the approach proposed in the Colombo project does not provide incentives for unequipped vehicles to install DSRC radios as well. Hence, it is not clear if with such an approach the system can scale easily. This appears to be a major limitation of the Colombo project.

Meanwhile, an infrastructure free intersection coordination scheme, known as Virtual Traffic Lights (VTL) has been introduced as a viable alternative solution to traffic management at intersections \cite{ferreira2010self}.
VTL uses Vehicle-to-Vehicle (V2V) Communications to manage the traffic at an intersection in a self-organized manner, thus,  the right-of-way is decided in a distributed manner. Extensive simulations have shown that VTL technology can reduce daily commute time of urban workers by more than 30\%. Different aspects of VTL technology, including algorithm design, system simulation, deployment policy, and carbon emission have been studied by different research groups in the last few years.\cite{ferreira2010self,neudecker2012feasibility,ferreira2012impact,nakamurakare2013prototype,viriyasitavat2013accelerating,bazzi2014distributed,hagenauer2014advanced,tonguz2014implementing,yapp2015safety,bazzi2016distributed,tonguz2016self}. VTL has the advantage over infrastructure based schemes mentioned earlier in that it is a much lower cost solution. However, VTL also assumes 100\% penetration of DSRC technology which will happen over time. In the meantime, a cost-effective transition scheme between current traffic control systems and VTL is needed.

The main contribution of this paper are:
\begin{enumerate}
  \item Report such a new scheme, DSRC-ATL, that works under low
penetration rates (i.e., when a small ratio of vehicles, such as 20\%, are equipped with DSRC technology). The scheme can be treated as both a low cost alternative to traditional ITS and a viable transition scheme for VTL.
  \item Report a prototype implementing the scheme proposed. The prototype is evaluated in the field to test the communication level performance and simulations are performed to evaluate the overall performance in waiting time.
  \item The simulations are performed to evaluate the performance of DSRC-equipped and unequipped vehicles separately, hence, they illustrate how the scheme will help the transition process of DSRC radios to adopt larger penetration rate. To the best of our knowledge, none of the previous studies reported address this important practical problem.

\end{enumerate}

\section{DSRC-Actuated Traffic Lights System Design}
The DSRC-Actuated Traffic Light System we introduced in this paper has an "On Roadside" unit that communicate with the DSRC radios which already start to be installed on vehicles in USA. In the next section we describe in detail the components and their functions. Notice that any standard DSRC radio on the vehicle will be compatible with the system we design, no extra configuration, installation or software is needed on the vehicle side.

\subsection{Components of the  DSRC-ATL Prototype System}

\begin{figure}[ht]
\centering

\includegraphics[width=3in]{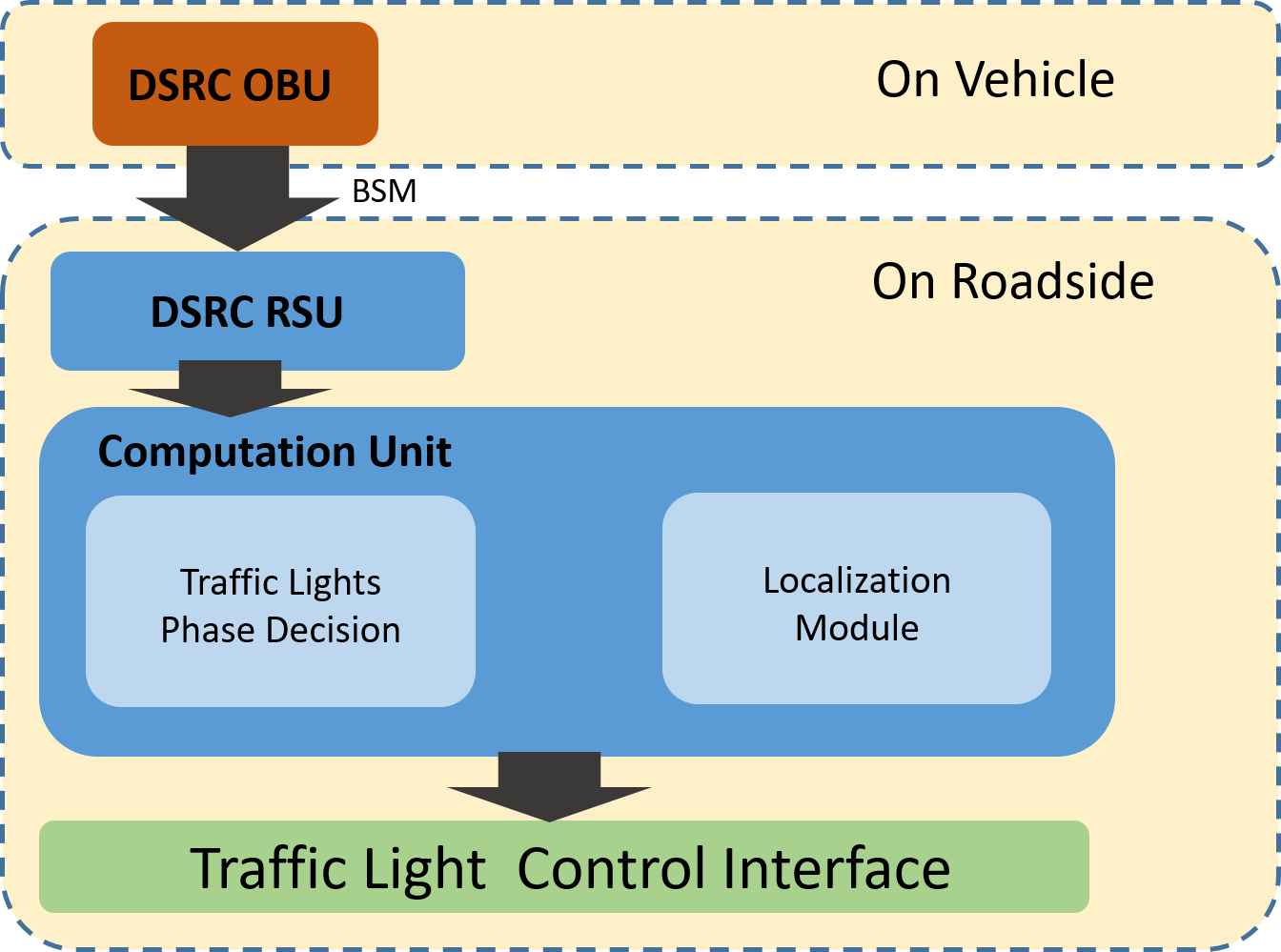}
\caption{Components of the  DSRC-ATL prototype system.}
\label{fig_components}
\vspace{-0.1in}
\end{figure}

Figure \ref{fig_components} shows the block diagram of DSRC-ATL prototype system, with  "On Vehicle" and "On Roadside" blocks. "On Vehicle" has the standard DSRC On Board Unit (OBU) that newer vehicles come equipped with. "On Roadside" block is made up of three main functional blocks: DSRC RoadSide Unit (RSU), Computational unit and Traffic Light Control Interface unit. DSRC RSU receives Basic Safety Messages (BSM) transmitted from the DSRC OBU on the vehicle. The received messages are then passed to the Computation Unit. The Computation Unit processes the messages received and runs traffic signal control algorithm based on the received messages and the phases of traffic lights are maintained accordingly.

\subsubsection{DSRC OBU and RSU}
\label{SS: radio}

\begin{figure}[!ht]
\centering
\subfloat[DSRC RSU]{\includegraphics[width=1.7in]{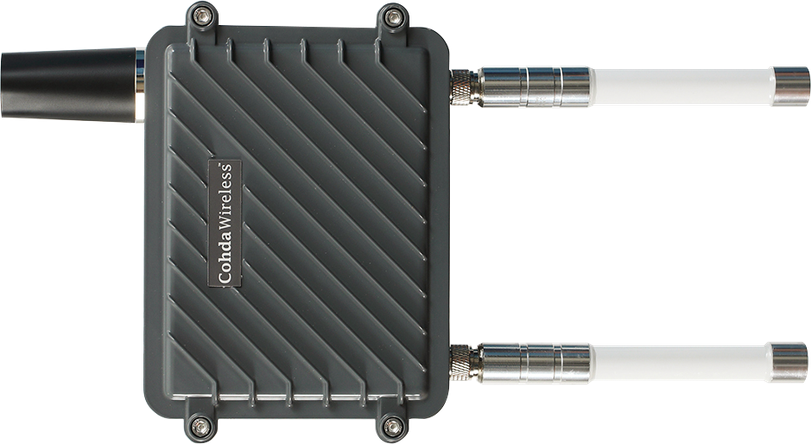}%
\label{fig:DSRC_RSU}}
\hfil
\subfloat[DSRC OBU]{\includegraphics[width=1.3in]{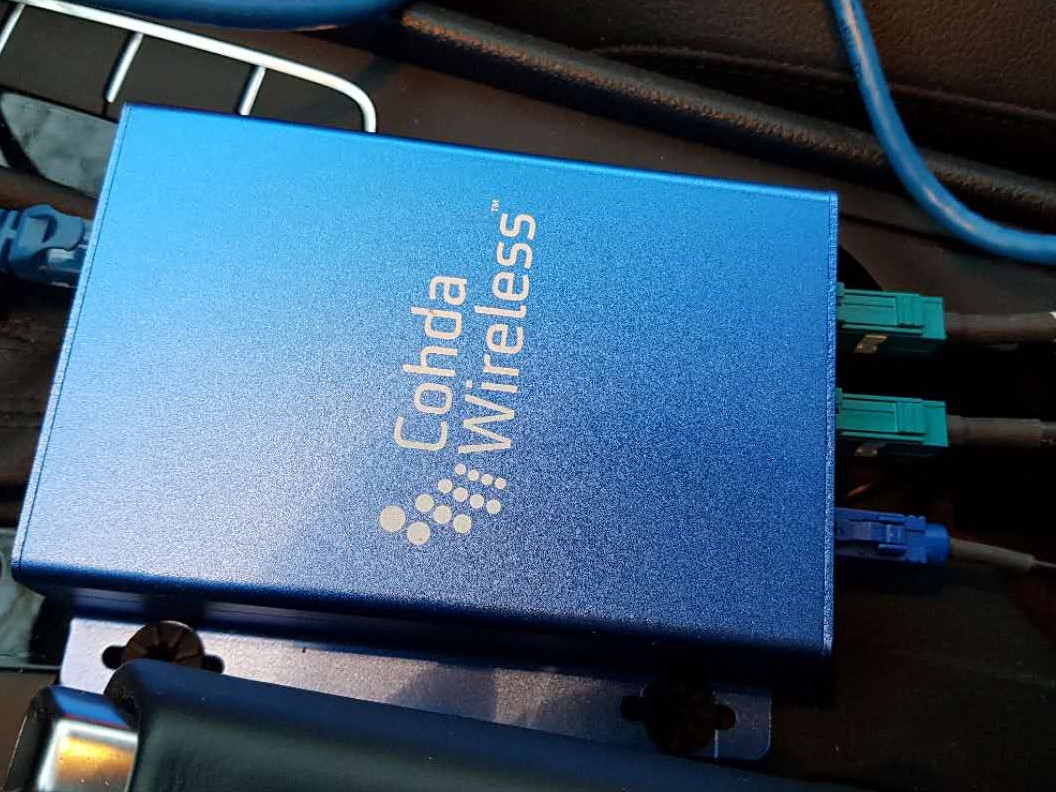}%
\label{fig:DSRC_OBU}}
\caption{DSRC radios used in the prototype system}
\label{fig_DSRC}
\vspace{-0.05in}
\end{figure}

DSRC radio is a short to medium range radio working in the 5.9 GHz band (5.850-5.925 GHz), which is allocated by the Federal Communications Commission (FCC) to be used for vehicle-related safety and mobility systems \cite{what, kenney2011dedicated}. A DSRC radio can be working inside of a vehicle as an OBU or at road-side as a RSU. Figure \ref{fig_DSRC} shows the DSRC radio used in our prototype system, manufactured by Cohda Wireless.


The DSRC OBU is capable of broadcasting several types of messages specified by the SAE 2735 protocol \cite{sae20142735}. BSM is the most important  message type to be broadcast by each OBU at a certain frequency (normally 10 Hz) and it provides situational data to its surroundings. BSM contains a vehicle's current information, including its GPS coordinates, speed,  heading, and a temporary ID. By sensing the BSM broadcast by the vehicles, an RSU is able to detect vehicles coming toward the intersection. Traditional detection methods, such as loop detectors, only detect presence of vehicles, while DSRC detects vehicles in a continuous manner.

The DSRC RSU is the device to facilitate the communication between vehicles and transportation infrastructures, in our case the traffic lights. The RSU picks up BSMs broadcast by vehicles, filters and parses the information from BSMs and transmits them to the Computation Unit for further processing. Since the BSM is the only type of messages the system needs, the system does not require any software or any personalized configuration on the OBU.

\subsubsection{Computation Unit}
Computation Unit takes information from DSRC RSU and decides the right-of-way for all approaches of the intersection.  It then transmits its decision through Traffic Light Control Interface to Traffic Light Control Box (TCB) and actuate phase of the traffic lights accordingly.

The application running on the computation unit is composed of two modules: a Traffic Lights Phase Decision Module and a Localization Module. The Localization Module takes the GPS coordinates and provides geo-information, needed for the Traffic Lights Phase Decision Module to make decisions.

\begin{figure}[htb]
\centering
\includegraphics[width=3in]{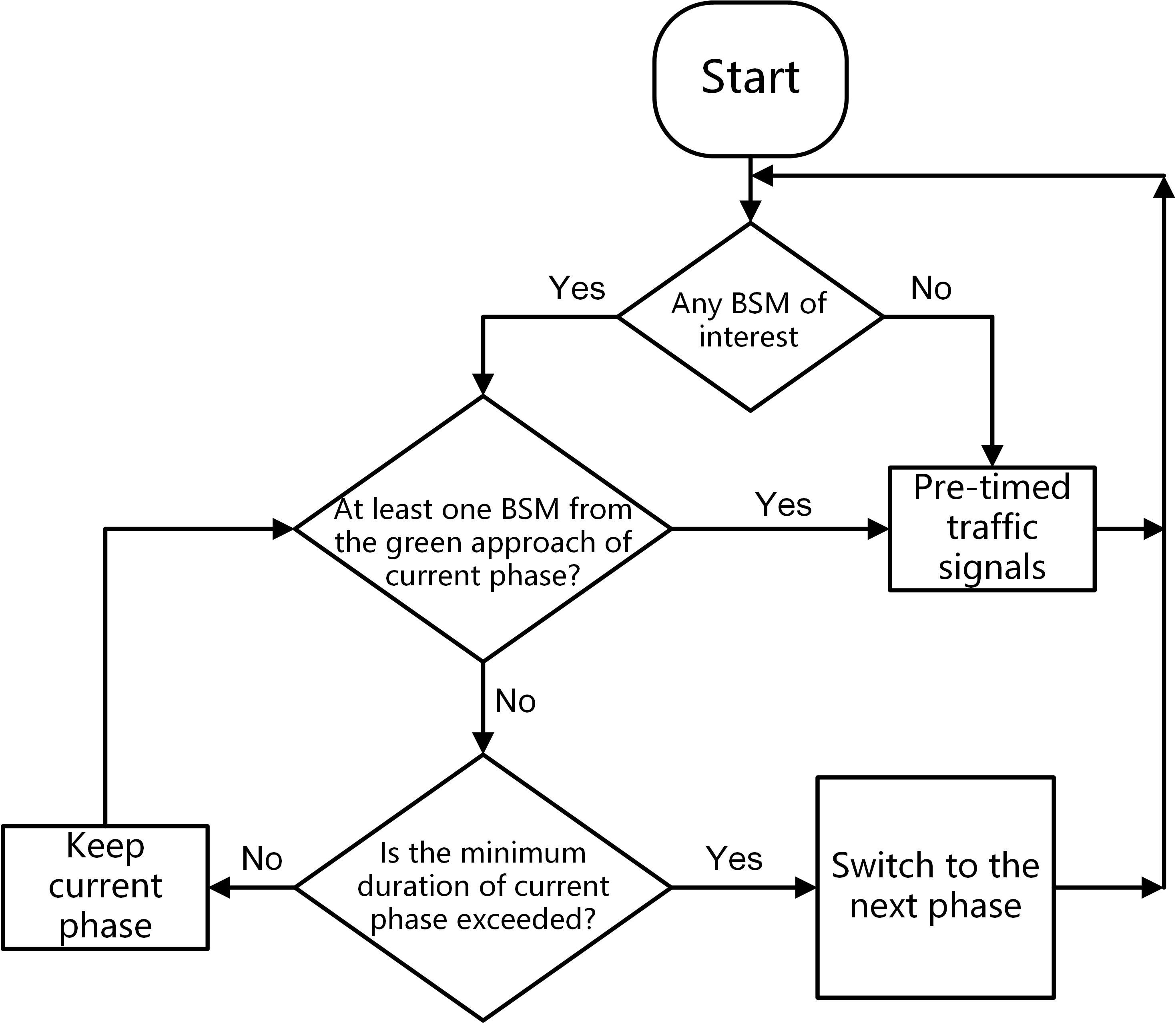}
\caption{Traffic Lights Phase Decision logic.}
\label{fig_phase_decision_logic}
\vspace{-0.1in}
\end{figure}

Figure \ref{fig_phase_decision_logic} shows the overall principle of operation of the Phase Decision Algorithm as a flow chart. Observe that the RSU is continuously checking if BSM from the area of interest is received. If no BSM is received, then the scheme regresses to pre-timed traffic signal. If the system detects the presence of BSMs, then it checks whether the detected BSMs are from the approaches that currently have the green light. If so, then the algorithm moves to the pre-timed operation mode. If not, then this implies that the DSRC-equipped vehicles are in an approach that currently has the red phase. In this case, the system checks whether the current time that has lapsed is larger than the minimum time allowed for the green phase. If so, then switching occurs and the approach that includes the DSRC-equipped vehicles gets the green light.

\subsubsection{Traffic Control Box}

Traffic Light Control Box (TCB) systems are installed for controlling vehicle flows in an intersection efficiently. A typical TCB has the ability to control traffic in a fixed time or an adaptive mode. Fixed time controllers allow a predefined fixed time for each lane while adaptive controllers take traffic information such as vehicle density into consideration. In order for adaptive controllers to work properly it needs to interface with traffic detectors such as induction loops, cameras, pedestrians and OBUs. Standard such as ISO 10711:2012 \cite{ISO10711} defines such interfaces between controllers and detectors. Additionally, National Transportation Communication for ITS Protocol (NTCIP) defines various interfaces requirements for traffic lights systems' manufacturer in the USA \cite{national2002ntcip}.

\begin{figure}[!ht]
\centering
\subfloat[Traffic control box]{\includegraphics[width=1.2in]{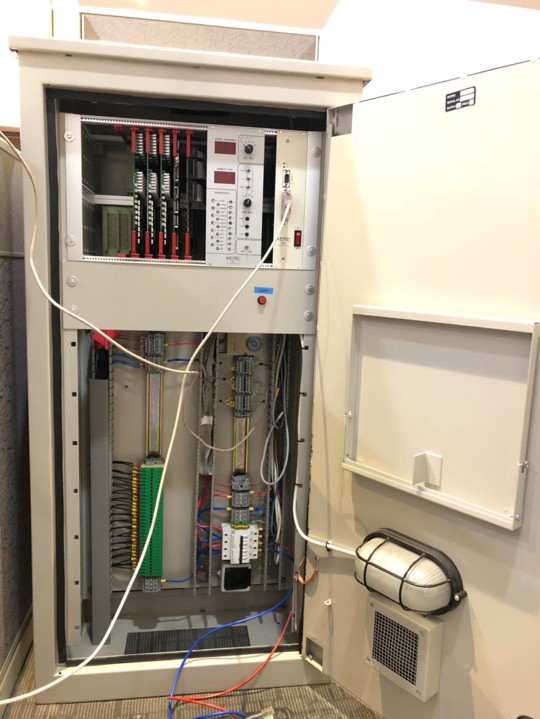}%
\label{fig:TCB}}
\hfil
\subfloat[Controller in the traffic control box]{\includegraphics[width=1.8in]{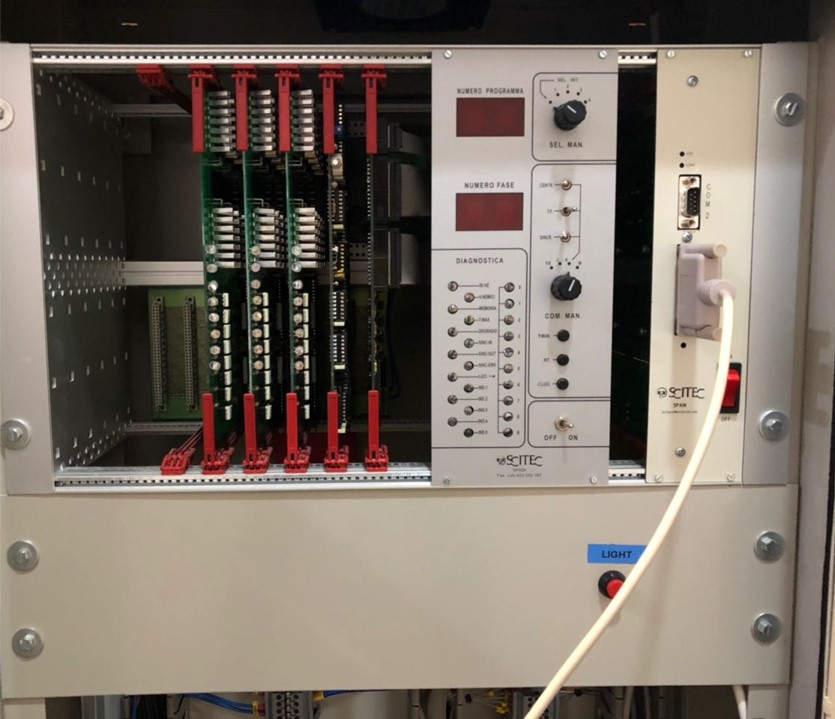}%
\label{fig:TCB_controller}}
\caption{Traffic control box used for the prototype}
\label{fig_TCB}
\vspace{-0.05in}
\end{figure}

Figure \ref{fig_TCB} shows the TCB we use for the prototype. The TCB is interfaced with the computation unit and all TCB phases are controlled and activated adaptively based on the output of Traffic Lights Phase Decision module. In our prototype, the phases were programmed in the TCB plan with each phase configured to be activated when a specific input is triggered. Simple relays were used in the physical connection between the computation unit and those inputs in order to protect the computation unit circuit from high voltage.

\section{Performance}
In this section, we report some promising  field test results and simulation results using the system introduced above. We present communication level results that show the communications performance between vehicles and RSU at the intersection. We also present system level results that show the end-user benefits.
\subsection{Communication Performance}
\label{ss:comm_level}

\begin{figure}[htb]
\centering
\includegraphics[width=3in]{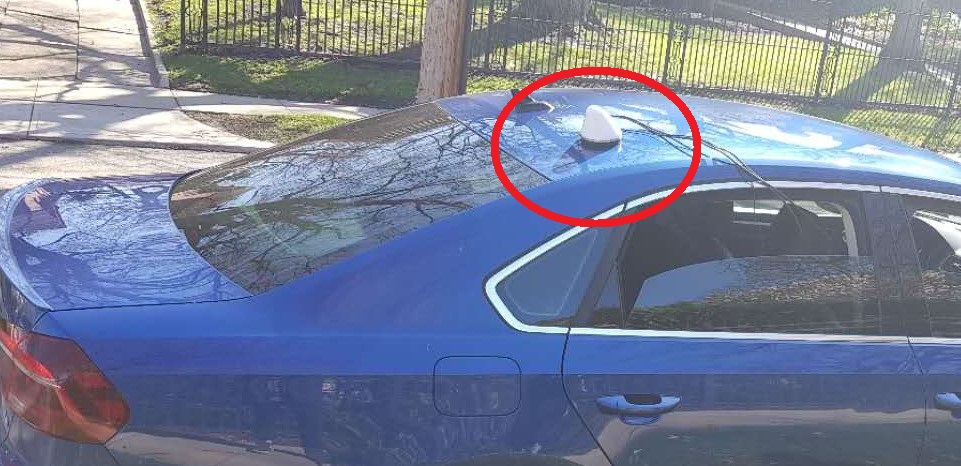}
\caption{Outside view of the vehicle, showing the DSRC antenna, circled.}
\label{fig_antenna}
\vspace{-0.1in}
\end{figure}

\begin{figure}[!ht]
\centering
\subfloat[The intersection where the field trial is running at Ellsworth and Amberson avenues, Pittsburgh, PA.]{\includegraphics[width=1.8in]{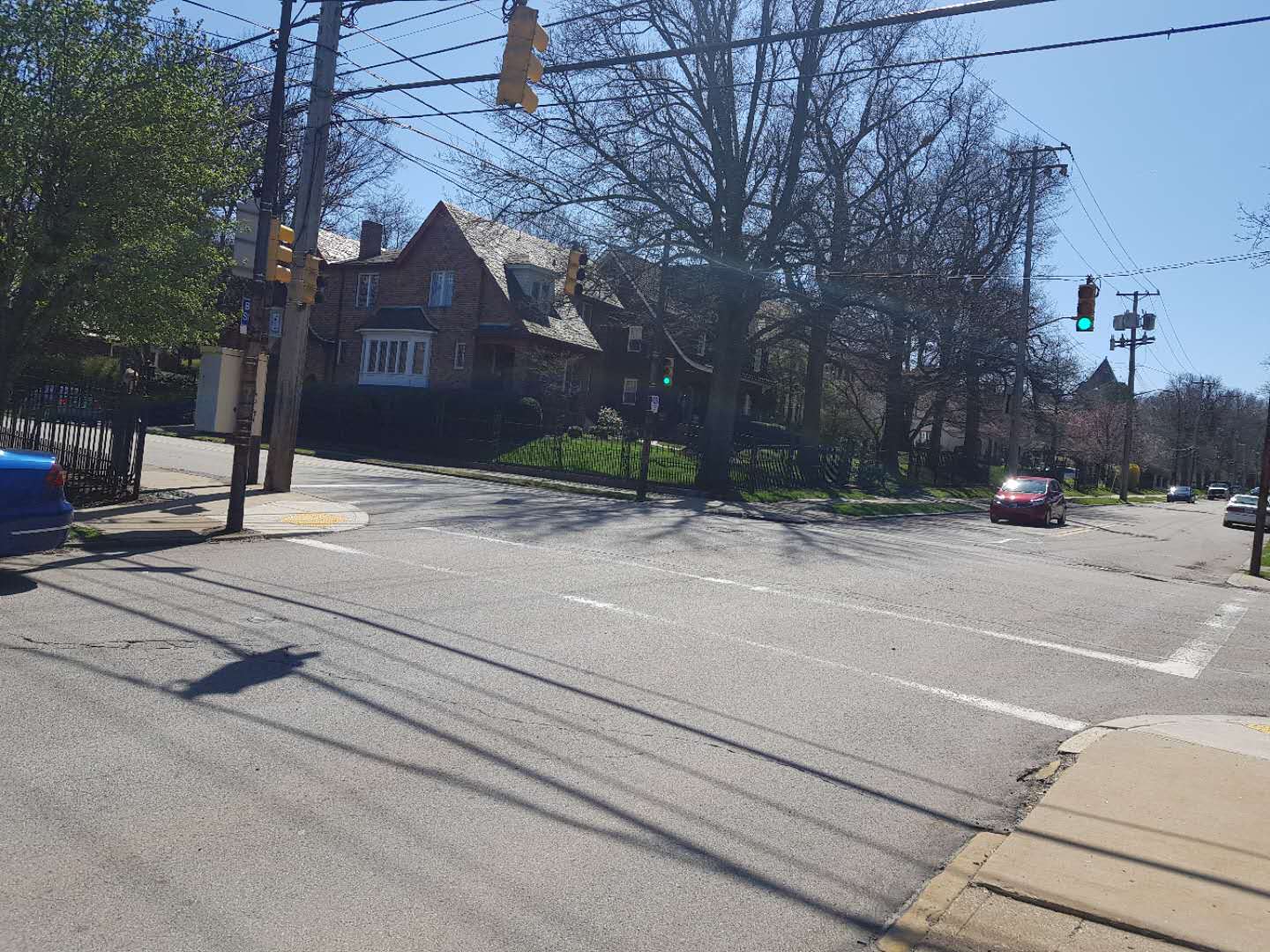}%
\label{fig:intersection}}
\hfil
\subfloat[RSU is installed 2.5 meters above the ground, circled in the figure]{\includegraphics[width=1.2in]{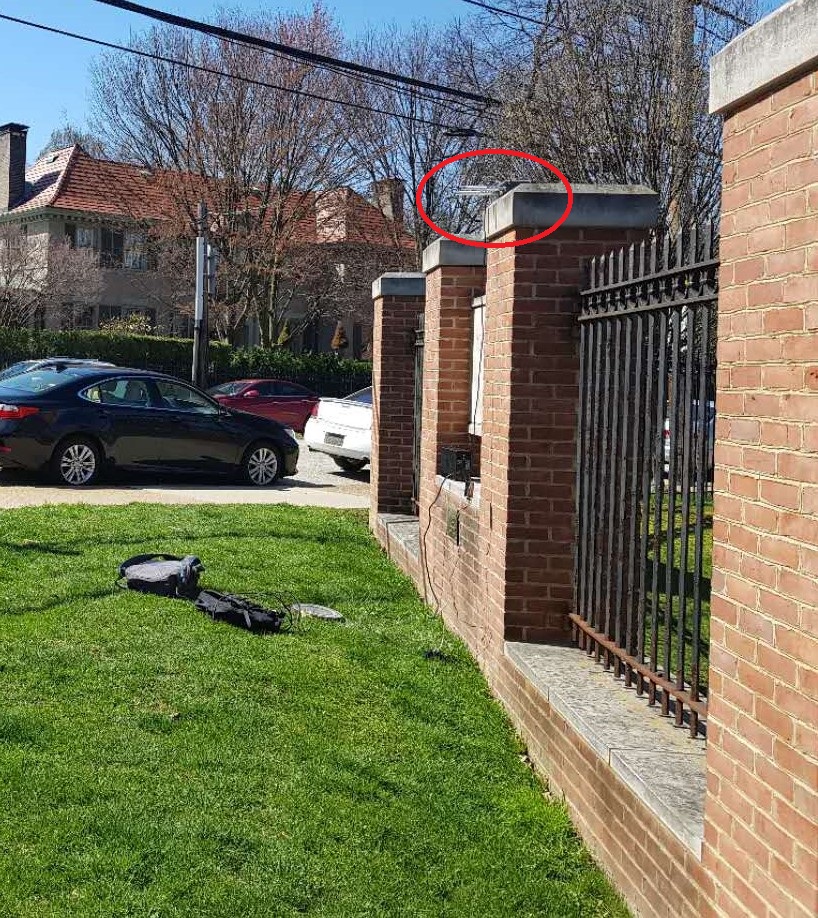}%
\label{fig:mount}}
\caption{Intersection of the field trial.}
\label{fig_int}
\vspace{-0.1in}
\end{figure}

The communications between DSRC radios is the key factor that enables the whole functionality of the system. While there are multiple metrics to evaluate communications performance, in our application, the most important metric is Inter-Packet Gap (IPG), which is the time duration between two successfully received packets. In the context of this paper, IPG is the duration between two successfully received BSMs. Since RSU is using the received BSMs to determine other vehicles' situation, IPG directly determines how continuously the vehicle is sensed by the RSU.

In our experiment, we mount the RSU 2.5 meters above the ground, as shown in figure \ref{fig:mount}. Vehicle transmits BSM at a frequency of 10 Hz. We record the IPG of BSM received at RSU when the vehicle is in distance from 25 meters to 150 meters toward the intersection. There is a test point every 25 meters. The vehicle is heading to the intersection, and the DSRC OBU antenna is at the rear end of the vehicle, where the antenna of the vehicle is typically situated, as shown in Figure \ref{fig_antenna}. At each test point, more than 300 BSMs are transmitted, and we report the average IPG from different distances to the intersection.

\begin{figure}[hbt]
\centering
\includegraphics[width=3.5in]{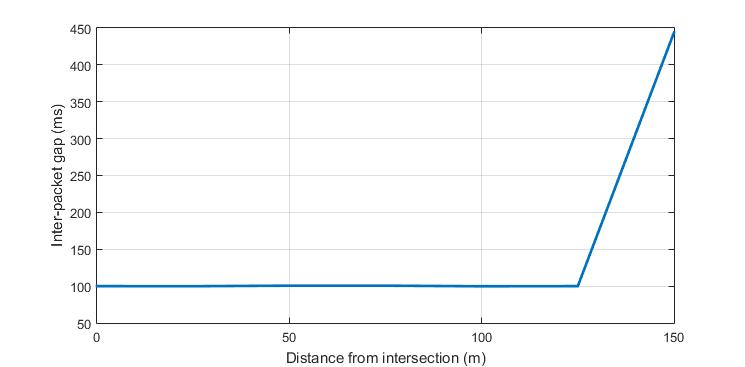}
\caption{Relations between Inter-Packet Gap (IPG) and distance
}
\label{fig_results}
\vspace{-0.1in}
\end{figure}

Figure \ref{fig_int} shows a picture of the intersection in Pittsburgh where we collect data. Observe that there are houses and trees between the transmitter and receiver, and this corresponds to a Non-Line-Of-Sight (NLOS) scenario. Figure \ref{fig:mount} shows the specific part of the intersection where we mount the RSU (circled in the figure), 2.5 meters above the ground. In a commercial implementation, the RSU will be mounted higher, where the obstacles between RSU and OBU will be fewer, and thus the communications quality will be better. Therefore, the results obtained here reflect a close, yet lower bound of commercial system performance.

Figure \ref{fig_results} shows the relations between Inter-Packet Gap and distance from the intersection. Observe that as the distance from the intersection decreases, shown from 150m to 0m, the IPG has a steep decent up until 125m from the intersection, and it levels off at about 100 m from the intersection.
Since we only need to detect vehicles at around 50 meters from the intersection, this IPG performance is more than sufficient.
The excellent performance of DSRC radios when the distance is less that 125 meters indicates that the DSRC radios are fully capable of sensing vehicles approaching the intersection. 

\subsection{System Level Performance}
\label{ss:system-level}

To evaluate the performance of this system when it is in commercial deployment, we use  an open-source mobility simulator known as SUMO \cite{krajzewicz2012recent}.

\begin{figure}[hbt]
\centering
\includegraphics[width=3in]{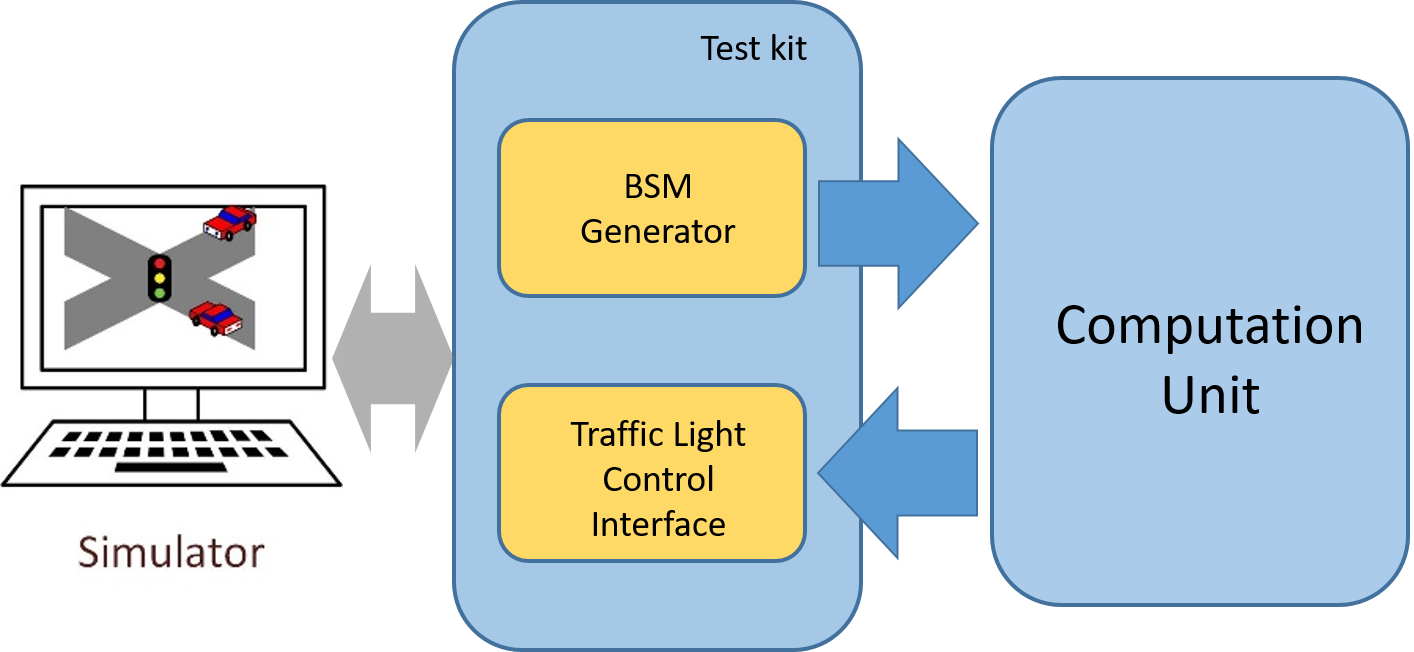}
\caption{Test kit for debugging and performance analysis.}
\label{fig_testkit}
\vspace{-0.1in}
\end{figure}

We developed a test kit based on SUMO to mimic the real world behavior and record the performance. Figure \ref{fig_testkit} shows the structure of the test kit.
The test kit's BSM generator will take vehicles' situation in the simulator, and generate exactly the same format of BSM and send to the same port of the computation unit. The computation unit process the BSM information and output the traffic command to the traffic light control interface of the test kit, which will take same type of the command as the real traffic light control interface. The traffic light control interface of the test kit then changes the traffic light's status in the SUMO simulator. Since the formating of the data, and the interface between the module are the same, the test kit will be able to predict the actual performance of the algorithm shown in Figure \ref{fig_phase_decision_logic}, as long as the communication between RSU and OBU are robust, which has been confirmed from the results reported in subsection \ref{ss:comm_level}. Hence, this test kit has an important role for debugging and performance analysis.

\begin{figure}
  \centering
  \includegraphics[width=0.9\columnwidth]{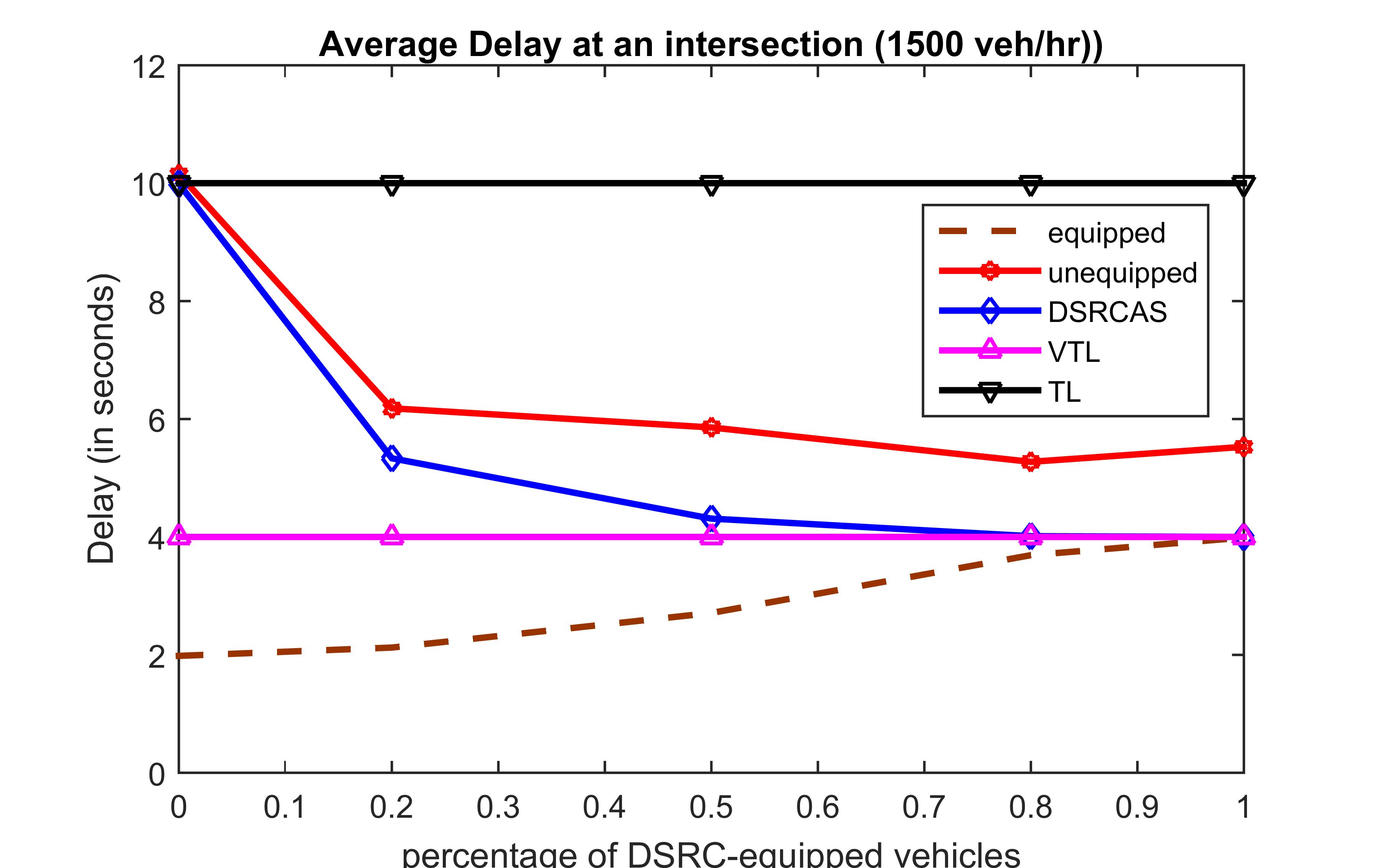}\\
  \caption{Predicted performance of the prototype at an intersection in terms of Average Waiting Time as a function of DSRC-equipped rate. For comparison, the average waiting time of regular traffic lights (TL) and Virtual Traffic Lights (VTL) is also shown.}\label{fig_performance}
\end{figure}

We present results with car flow of 1500 cars/hr, which is the average arrival rate of a one-lane road \cite{numbers}, the arrival pattern of the cars is assumed to be a Poisson arrival, which is typical in traffic engineering simulations, the ratio of car flow between the two avenue is assigned to 4:1, which is roughly the actual car flow ratio between Ellsworth Avenue and Amberson Avenue, where we do the field trial. The waiting time at the intersection is quantified. To provide a detailed analysis, the waiting time for DSRC-quipped and unequipped vehicles are given in addition to the overall system performance of prototype system. To put things in perspective, the performance of current traffic lights (TL) and VTL system are also provided that allows a more meaningful comparison which, in turn, leads to a better understanding of the benefits of the proposed system as a function of the DSRC-equipped rate.

Figure \ref{fig_performance} shows the results. The overall system performance of the DSRC-actuated traffic light (DSRC-ATL) asymptotically approaches the performance of VTL system. An interesting observation that can be made from Figure \ref{fig_performance} is the fact that a large portion of the improvement with the DSRC-Actuated Traffic Control scheme occurs with modest levels of penetration (about 80\% of the total improvement occurs when only 20\% percent of vehicles are equipped with DSRC radios) which is a very interesting and attractive feature. In other words, with a relatively modest penetration rate of 20\%, one gets a huge improvement with respect to the TL scheme.

Another quite interesting fact is that, the waiting time of DSRC-equipped vehicles is shorter than the waiting time of unequipped vehicles during the whole transition process, which provides a compelling reason and motivation for end-users to install DSRC radios in vehicles other than all the safety applications already implemented.

Furthermore, when one reaches 60-70\% penetration ratio, if DSRC radios in certain vehicles malfunction or they get out of traffic stream, the degradation experienced is almost negligible . This shows the robustness of the proposed scheme which is again a very desirable feature.

\section{Conclusion}
In this paper, we report a new prototype system known as DSRC-actuated traffic lights, which is a DSRC based intelligent traffic system that can work under a low penetration ratio of DSRC-equipped vehicles. In addition to presenting a detailed system and prototype design, we also report the results of extensive field trials carried out in Pittsburgh.

Our field trial results show that the RSU designed as part of our prototyping effort can indeed sense vehicles continuously when vehicles are more than 100 meters away from the intersection, which provides strong evidence about the viability of a new sensing technology for intelligent intersections.

We have also performed a performance analysis based on the test kit we developed. The results provide compelling evidence that the designed prototype and system can reduce the average delay at an intersection even when the vehicles equipped with DSRC radios are only a small percentage of the overall vehicles using that intersection. Our current work is focused on extending these results to several intersections on an arterial road in Pittsburgh.







\bibliographystyle{IEEEtran}
\vspace{-0.05in}
\bibliography{IEEEabrv,reference}
%



\end{document}